\documentclass[prl,twocolumn,showpacs,preprintnumbers]{revtex4}
\usepackage{amsmath,amssymb,epsfig}
\begin{document}

\title{Exact many-electron ground states on the diamond Hubbard chain}
\author{Zsolt~Gul\'acsi$^{a,b}$, Arno~Kampf$^{a}$, and
Dieter~Vollhardt$^{a}$}
\address{$^{(a)}$ Theoretical Physics III, Center for
Electronic Correlations and Magnetism, Institute for Physics,
University of Augsburg, D-86135 Augsburg, Germany \\
$^{(b)}$ Department of Theoretical Physics, University of
Debrecen, H-4010 Debrecen, Hungary}
\date{March 30, 2007}
\begin{abstract}
Exact ground states of interacting electrons on the diamond Hubbard
chain in a magnetic field are constructed which exhibit a wide range
of properties such as flat-band ferromagnetism and correlation
induced metallic, half-metallic or insulating behavior. The
properties of these ground states can be tuned by changing the
magnetic flux, local potentials, or electron density.
\end{abstract}
\maketitle

Condensed matter systems with macroscopic degeneracies react very
sensitively on internal or external perturbations and thus give rise
to fascinating emergent behavior. Well-known examples are electrons in a
magnetic field in two dimensions \cite{QHE} and spins on lattices with
geometric frustration \cite{Ramirez}. Dispersionless ("flat")
electronic bands in solids also lead to macroscopic degeneracies.
Recent advances in nanotechnology allow for the possibility to
design simple structures, which have flat \emph{single-electron}
bands, i.e., when electronic correlations are neglected
\cite{int1,int1a,int2,int5}. The controlled setup of optical
lattices for cold atoms using standing wave laser light-fields also
allows to realize a variety of lattice models of interacting
fermions and bosons \cite{Jaksch,Wu}. The ability to generate flat
bands by changing external parameters such as a gate voltage or a
magnetic field, would make even a direct manipulation of
macroscopically degenerate systems possible. For
example, the understanding of ferromagnetism in organic compounds
\cite{int2} could be improved, and the tuning of instabilities or
the switching between different phases would permit direct
applications in spintronics.

Most exact flat-band results concern flat \emph{lowest}-energy
bands, and provide solutions for ground states or the low
temperature thermodynamics \cite{Wu,Derzhko}. For such a case  Mielke
and Tasaki proved that ferromagnetism is stabilized at or near
half-filling \cite{int6}.
Lieb's ferrimagnetism emerges on a bipartite lattice with a
macroscopically degenerate energy level exactly in the middle of the
spectrum \cite{int7}. For the more general case, when the
interaction leads to an effective flat band above a dispersive band,
exact results were derived for the periodic Anderson model
\cite{zv}, by exact diagonalization of small clusters and \emph{ab
initio} band-structure calculations \cite{int2}, or analytically for
the two-particle problem \cite{int5}.

In this Letter we concentrate on one of the simplest lattice
electron models, the Hubbard model on a diamond chain, as
found, e.g., in azurite. In the diamond chain flat bands occur
already in the one-electron picture. External potentials or a 
magnetic field can give rise to additional flat bands. For selected 
parameter sets, and by appropriately tuning local potentials or the 
magnetic flux, we construct exact many-body ground states
on the diamond chain, which are either insulating or conducting and
fully or partially spin polarized. Our results thus open new routes
for the design of spin valve devices and gate induced
ferromagnetism.

\begin{figure}[b!]
\epsfxsize=7.55cm \centerline{\epsfbox{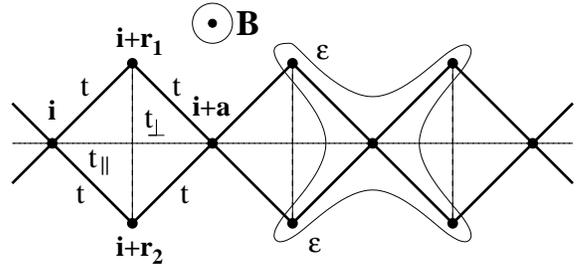}}
\caption{Diamond Hubbard chain; the cross-shaped region depicts a
localized (Wannier) eigenfunction for $t_\perp=t_\parallel=0$ and
flux $\delta =\pi/2$ (see text). } \label{fig1}
\end{figure}

Fig. 1 shows the diamond chain, whose sites are denoted
by ${\bf i}+{\bf r}_s$, where ${\bf i}$ and ${\bf r}_s$ (with
sublattice index $s=1,2,3$ and ${\bf r}_3={\bf 0}$) denote the
unit cells and the three sites inside, respectively. The
Bravais vector of the chain is ${\bf a}$; periodic boundary
conditions along the chain are assumed. Given $N$ electrons
their density is $n=N/(3N_c)$, where $N_c$ is the number of
unit cells. The Hamiltonian for the diamond Hubbard chain has the form
\begin{eqnarray}
{\hat H}=\sum_{{\bf k},\sigma}\sum_{s,s'=1}^3 M_{s,s'}({\bf k})\, \hat
c^{\dagger}_{s,{\bf k},\sigma} \hat c_{s',{\bf k},\sigma} + {\hat
H}_{U}\, ,\nonumber\\
{\hat H}_{U}=U\sum_{{\bf i}}\sum_{s=1,2,3}\hat n_{{\bf i}+{\bf r}_s,
\uparrow} \hat n_{{\bf i}+{\bf r}_s,\downarrow} \, , \label{E1}
\end{eqnarray}
where the kinetic and the interaction part are written in $\bf k$
space and position space, respectively. Here and in the following
sums or products over $\bf k$ and $\bf i$ extend over the $N_c$
vectors enumerating the unit cells, $\hat c^{\dagger}_{{\bf
j},\sigma}$ creates an electron with spin $\sigma$ at site ${\bf
j}$, the local density is given by $n_{{\bf j},\sigma}={\hat
c}^\dagger_{{\bf j}, \sigma}{\hat c}_{{\bf j},\sigma}$, and $\hat
c_{s,{\bf k},\sigma}$ is the Fourier transformed sublattice
operator. The elements of the symmetric matrix $M_{s,s'}({\bf k})$
in the kinetic energy are
\begin{eqnarray}
&&M_{1,1}=M_{2,2}=\epsilon, \quad M_{3,3}=2t_{\parallel} \cos ak,
\quad M_{1,2}=t_{\perp},
\nonumber\\
&& M_{(s=1,2),3}= 2t \cos [(a k+(-1)^s \delta)/2] . \label{Mmatrix}
\end{eqnarray}
Here, $t$, $t_{\perp}$, $t_{\parallel}$ denote the hopping matrix
elements to nearest neighbors (NN) and to next NN sites
(perpendicular and parallel to ${\bf a}$), respectively. A Peierls
phase factor $\exp({\rm i}\delta/2)$ with $\delta=2\pi\Phi/\Phi_0$
accounts for a perpendicular magnetic field ${\bf B}$;
$\Phi$ is the flux threading the unit cell and $\Phi_0=hc/e$ is the
flux quantum. Choosing the vector potential ${\bf A}\parallel{\bf
a}$ in the field dependent hopping amplitudes $t_{{\bf j},{\bf
j}'}({\bf B})=t_{{\bf j}, {\bf j}'}(0)\exp[({\rm
i}2\pi/\Phi_0)\int_{\bf j}^{{\bf j}'}{\bf A}\cdot d {\bf l}]$ the
magnetic field does not alter $t_{\perp}$ and $t_{\parallel}$. The
on-site potential $\epsilon$ acts on the sublattices $s=1,2$.
$U$ is the on-site Hubbard repulsion. In
the following energies will be in units of $2t$.

The diagonalization of the kinetic energy part in Eq. (\ref{E1})
leads to a cubic eigenvalue equation for the band dispersion $E({\bf
k})$. With $P_1=(1+t_\perp^2)+\cos\delta\cos ak$ and $P_0=t_\perp [
\cos\delta +\cos ak+ t_\perp (\epsilon -2 t_{\parallel} \cos ak )]$
the dispersion $E_\nu({\bf k})$ ($\nu=1,2,3$) is determined by
\begin{equation}
(E-\epsilon)^3+(E-\epsilon)^2 (\epsilon-2t_{\parallel}\cos ak)
- (E-\epsilon) P_1 = P_0 .
\label{E4}
\end{equation}
Depending on the choice of $t_\perp$, $t_\parallel$, the magnetic
field, and the potential $\epsilon$, the single-electron band
structure contains one, two or even three exactly flat bands
(Fig. 2). It should be stressed that these bands lose their meaning
as soon as the interaction is switched on.

\begin{figure}[t!]
\epsfxsize=7.5cm \centerline{\epsfbox{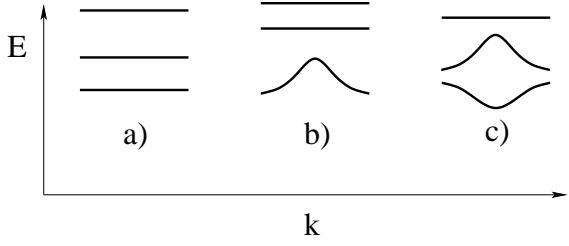}}
\caption{Single-electron band structure: (a)
$t_{\perp}=t_{\parallel}=0$, $\delta= \pi/2$. (b)
${\epsilon}=-t_{\perp}+t_{\perp}^{-1}/2$, $\delta=\pi$; the upper
flat bands are degenerate for $t_{\perp}t_{\parallel}=1/4$. (c)
$t_{\parallel}=0$, $\epsilon t_{\perp}\cos\delta=t_{\perp}^2-
\cos^2\delta$.} \label{fig2}
\end{figure}

Vidal \emph{et al.} recently presented a detailed study
of two electrons on the diamond Hubbard chain in the limit
$t_\perp=t_\parallel=\epsilon=0$ \cite{int5}.
They showed that for half a flux quantum per unit square
($\delta=\pi/2$) the excited singlet eigenstates are
localized if $U=0$, but become delocalized for $U>0$. Apparently,
the interaction  $U$ is able to induce subtle correlations leading
to conducting states, which led them \cite{int5} to speculate that
such a delocalization also holds for a finite electron density.

Here we construct exact many-particle ground states for quite
general cases and thereby also resolve some of the
issues raised in \cite{int5}. The strategy for deducing exact ground
states was described before in the context of the periodic
Anderson model \cite{zv}, but has not yet been applied to Hubbard
models. The key steps are to first cast the Hamiltonian in positive
semidefinite form and then to construct an explicit eigenstate with
minimal energy.

\emph{Solution I. Flat-band ferromagnetism:} We start with localized
ground states for densities $n\leq 1/3$, $t_\perp=t_\parallel=0$
and $\delta =\pi/2$ ("Aharonov-Bohm cage" \cite{int5}), in which
case Eq. (\ref{E4}) provides three  flat single-electron bands
with energies $E_2=\epsilon$, $E_{2\pm 1}=(\epsilon
\mp \sqrt{{\epsilon}^2 +4})/2$ (see Fig. 2a). Introducing the
canonical fermionic operators $\hat C_{\nu,{\bf i},\sigma}$
\begin{equation}
\hat C_{2\pm 1,{\bf i},\sigma}={1\over\sqrt{2}}[{F_{\pm}} {\hat
Q}^{(+)}_{{\bf i}\sigma}\mp 2F_{\mp} \hat c_{{\bf i},\sigma}] \hskip0.1cm
,\hskip0.1cm \hat C_{2,{\bf i}, \sigma}={\hat Q^{(-)}_{{\bf i}\sigma}\over
2}, \label{Cop}
\end{equation}
with $F_{\pm}^2=1\mp {\epsilon}/\sqrt{{\epsilon}^2+4}$, $\hat
Q^{(\pm)}_{{\bf i}\sigma}=\hat Q^{(1)}_{{\bf i}\sigma}(\mp\delta)\pm\hat
Q^{(2)}_{{\bf i}\sigma}(\pm \delta)$ and
$\hat Q^{(l=1,2)}_{\bf i}(\delta) = {\rm e}^{-{\rm i}\delta/2}\hat
c_{{\bf i}+{\bf r}_l,\sigma} + {\rm e}^{+{\rm i}\delta/2}\hat
c_{{\bf i}-{\bf a}+{\bf r}_l,\sigma}$,
the Hamiltonian can be written in the form
\begin{equation}
\hat H=\sum_{{\bf i},\sigma}\sum_{\nu=1}^3 E_\nu \hat
C^{\dagger}_{\nu,{\bf i},\sigma}\hat C_{\nu,{\bf i},\sigma}+\hat
H_{U}. \label{E5}
\end{equation}
Since $\hat C^{\dagger}_{\nu,{\bf i},\sigma}\hat
C_{\nu,{\bf i},\sigma}$ and $\hat H_U$ in Eq. (\ref{E5}) are positive
semidefinite operators, the ground state of $N \leq N_c$ electrons at $U>0$ is
obtained by filling up the
eigenstates as
\begin{eqnarray}
|\Psi_{g}^I(N)\rangle = \prod_{{\bf i}=1}^N \hat C^{\dagger}_{3,{\bf
i}, \sigma_{\bf i}}|0\rangle; \label{E7}
\end{eqnarray}
$|0\rangle$ denotes the vacuum. The ground-state energy is
$E_{g}^{I}=E_3 N$. The operators $\hat C^{\dagger}_{\nu,{\bf
i},\sigma}$ create electrons in localized Wannier eigenstates; an
example of their structure (for $\nu=3$) is shown in Fig. (1).
For $n=1/3$ one obtains a unique, fully 
saturated ferromagnetic ground state. For $n<1/3$ only the Wannier 
states with a spatial overlap have the same spin. The
(degenerate) ground states hence consist of ferromagnetic clusters
of arbitrary spin orientation. When the density is increased to
$n=1/3$ the clusters touch and the degeneracy is lifted. Since $\hat
H_{U}|\Psi_{g}^I(N)\rangle=0$ and the kinetic part of $\hat H$ is
diagonal in real space, the ground state (\ref{E7}) is localized.
This is an explicit realization of Mielke-Tasaki's flat-band
ferromagnetism \cite{int6}. Since the lowest single-electron band is
flat only for $\delta=\pi/2$ and is dispersive for $\delta=0$,
and the system therefore most probably conducting, we here
encounter a metal-insulator transition as a function of magnetic
field.

\emph{Solution II. Correlated half-metal:} The results of Ref.
\cite{int5} suggest that itinerant states are easier to realize at
$\delta\ne\pi/2$. To investigate this point we analyze a group of
solutions for flux $\delta=\pi$, density $n\geq 4/3$, hopping
$t_{\perp}$, $t_{\parallel}>0$ and local potential
$\epsilon=1/(2t_{\perp})-t_{\perp}$. Then the bare band structure
consists of two upper flat bands $E_1,E_2$, and a lower dispersive
band $E_3({\bf k})$ (Fig. 2b). For simplicity we discuss here only
the case $t_{\perp}t_{\parallel}=1/4$, which implies $E_1=
E_2=\epsilon+t_{\perp}$ and $E_3=\epsilon-t_{\perp}- (1-\cos
ak)/(2t_{\perp})$. Electrons in the dispersive single-electron band
are created by the fermionic operators
\begin{equation}
\hat C_{{\bf k},\sigma}={\hat c_{2,{\bf k},\sigma}-\hat c_{1,{\bf
k}, \sigma}\over \sqrt{2} R_{\bf k}}+{\sqrt{1- \cos ak}\over 2t_{\perp}
R_{\bf k}}\, \hat c_{3,{\bf k},\sigma} \label{Ck3}
\end{equation}
with $R_k=\sqrt{1+(1-\cos ak)/(4t_{\perp}^2)}$. Defining
non-canonical fermionic operators \cite{non-canonical} as
\begin{equation}
\hat A_{{\bf i},\sigma}=\sqrt{t_{\parallel}}\, [\hat c_{{\bf
i},\sigma}-\hat c_{{\bf i}+{\bf a},\sigma}- 2 t_{\perp}
e^{i\frac{\delta}{2}} (\hat c_{{\bf i}+{\bf r}_1, \sigma} - \hat
c_{{\bf i}+{\bf r}_2,\sigma})] \,  \label{Aops}
\end{equation}
Eq. (\ref{E1}) is transformed into positive semidefinite form as
\begin{equation}
\hat H = \sum_{{\bf i},\sigma}\hat A_{{\bf i},\sigma} \hat
A^{\dagger}_{ {\bf i},\sigma} + U \hat P + E_{g}^{II};
\label{s4}
\end{equation}
$E_{g}^{II}=(\epsilon+U+t_{\perp})N-N_c[3U+4 t_{\perp}+1/t_{\perp}]$
is the ground-state energy. For the positive semidefinite operator
$\hat P=\sum_{{\bf j}=1}^{3N_c}(\hat n_{{\bf j},\uparrow}\hat n_{{\bf j},
\downarrow}-\hat n_{{\bf j},\uparrow}-\hat n_{{\bf j},\downarrow}+1)$
in Eq. (\ref{s4}) to assume its minimum eigenvalue $0$ there must be
at least one electron at each site. For $N=4N_c$ electrons ($n=4/3$) the
ground state then has the form
\noindent
\newcounter{equ}
\setcounter{equ}{1}
\def\theequation{\arabic{equation}.\alph{equ}}
\begin{eqnarray}
|\Psi_{g}^{II}(4N_c)\rangle &=&c\, [ \prod_{{\bf i}}
 \hat A^{\dagger}_{{\bf i},-\sigma}\hat A^{\dagger}_{{\bf i},\sigma}]\,
\hat F_{\sigma}^{\dagger}\, |0\rangle \\
\label{Eq1} \stepcounter{equ} \setcounter{equation}{10}
&=&\prod_{{\bf k}}\,[\hat C^{\dagger}_{{\bf k}, -\sigma}
\prod_{s=1}^3\hat c^{\dagger}_{s,{\bf k},\sigma}]\, |0\rangle ,
\label{Eq2}
\end{eqnarray}
\def\theequation{\arabic{equation}}
where $\hat F_{\sigma}^{\dagger}$ creates two electrons with spin
$\sigma$ on arbitrary sites of each unit cell; $c$ is a normalization
factor. Since $\hat A^{\dagger}_{{\bf i}, \sigma}$ creates one more spin
$\sigma$ electron in each unit cell, every lattice site is occupied by
a spin $\sigma$ electron. The electrons with spin $-\sigma$ are 
spatially extended, but they are localized in the
thermodynamic limit. This is inferred from the long-distance
behavior of the ground-state expectation value of the hopping term
$\Gamma_{\bf r,-\sigma} =\langle \hat c^{\dagger}_{ {\bf
j},-\sigma}\hat c_{{\bf j}+{\bf r},-\sigma}+H.c.\rangle$. With ${\bf
j}={\bf i}+{\bf r}_1$, $r/a=m$ and $N_c\rightarrow\infty$ one finds
$\Gamma_{m,-\sigma}=[(-1)^m\exp(-m/\xi_{-\sigma})]/\sqrt{1+1/t_{\perp}}$.
The one-particle localization length
\begin{eqnarray}
\xi_{-\sigma}= - \{ \ln [ 1+(2t_{\perp})^2-(2t_{\perp})
\sqrt{(2t_{\perp})^2+2}] \}^{-1}\, . \label{s4''}
\end{eqnarray}
is finite; it increases almost linearly with $1/t_{\perp}$.
For $N=4N_c+\Delta N$ electrons the ground state is given by
\begin{eqnarray}
|\Psi_{g}^{II}\rangle = \prod_{\alpha=1}^{\Delta N} \hat
c^{\dagger}_{n_{\alpha}, {\bf k}_{\alpha},-\sigma}
|\Psi_{g}^{II}(4N_c)\rangle,\label{s4'}
\end{eqnarray}
where $n_{\alpha}$ can take the values $s=1,2,3$.
Since the operators $\hat c^{\dagger}_{n_{\alpha}, {\bf
k}_{\alpha},-\sigma}$ add plane wave-type states to
$|\Psi_{g}^{II} \rangle$  the ground state now contains genuinely
 itinerant spin $-\sigma$ electrons \cite{localization_length}.
This, together with the fact that $\delta \mu (N)=0$
\cite{chemical_pot} for $1<\Delta N<N_c$ implies that the ground
state $|\Psi_{g}^{II} \rangle$ is conducting.
The net magnetization
decreases linearly with increasing electron density. $|\Psi_{g}^{II}
\rangle$ remains the ground state up to $\Delta N=N_c$ ($n=5/3$),
where it becomes non-magnetic. For densities $4/3<n<5/3$ the ground
state therefore corresponds to a \emph{correlation induced
half-metal}. Namely, while the $3N_c$ electrons with spin $\sigma$
are completely immobile and $N_c$ electrons with spin $-\sigma$ are
confined to their localized Wannier function, only the $\Delta N$
$-\sigma$ electrons are itinerant, leading to a low carrier-density
metallic behavior with a low spin polarization. Since the conduction
through this correlated half-metal involves only electrons of one
spin species, such a system may, in principle, serve as a spin valve
device, if contacted by metallic reservoirs \cite{Dieny,DFT}. Bearing
in mind that at $U=0$ the
electronic states at the Fermi level are dispersionless and hence
localized, we find that a finite, repulsive Hubbard interaction can
induce a localization-delocalization transition towards a
half-metal. This is an explicit example for the correlation-induced
conducting state conjectured by Vidal \emph{et al.} \cite{int5}. At
fixed magnetic field it can be realized by tuning the sublattice
potential $\epsilon$. Similar solutions at $\delta=\pi$ can be
deduced also at $t_{\parallel} t_{\perp}<1/4$, when the two upper
flat bands are non-degenerate.

\emph{Solution III.}
We now construct exact ground states for more general values of the
magnetic flux, $\delta\in (-\pi/2,+\pi/2)$, including zero flux,
for electron densities $n\geq5/3$. In particular, we will show that,
by switching on a magnetic field, a non-magnetic ground state may
turn into a non-saturated ferromagnet. We note that this occurs in
the absence of any Zeeman coupling to the spin and is only due to
the Peierls factor in the kinetic energy.
Specifically we select the parameters $t_{\parallel}=0$,
$t_{\perp}<0$, $b\equiv-\cos\delta/t_{\perp}$, $\epsilon =b-b^{-1}$.
In the non-interacting case one obtains a band structure with one
upper flat band at $E_1=\epsilon+1/b$, and two lower dispersive
bands at $E_{2,3}=\epsilon-b/2\pm (b^2/4+t^2_{\perp}+\cos\delta \cos
ak)^{1/2}$ (Fig. 2c). Defining the non-canonical fermionic operators
\cite{non-canonical}
\begin{equation}
\hat A_{\pm, {\bf i},\sigma}={1\over\sqrt{2b}}\left\{\begin{array}
{r@ {\quad}}{ b\hat
c_{{\bf i}+{\bf a},\sigma}-\hat c_{{\bf i}+{\bf r}_2, \sigma}
e^{{\rm }i{\delta\over 2}}-\hat c_{{\bf i}+{\bf r}_1,\sigma}
e^{-{\rm i}{\delta\over 2}}} \\b\hat c_{{\bf i},\sigma}-\hat c_{{\bf
i}+ {\bf r}_2,\sigma} e^{{-\rm }i{\delta\over 2}}-\hat c_{{\bf
i}+{\bf r}_1, \sigma}e^{{\rm i}{\delta\over 2}}\end{array}\right.
\label{Apm}
\end{equation}
the ground-state energy $E_{g}^{III}=(U+b)N-N_c(3U+4/b+2b)$ and the
Hamiltonian Eq. (\ref{E1}) in positive semidefinite form are obtained as
\begin{eqnarray}
\hat H=\sum_{\alpha=\pm}\sum_{{\bf i},\sigma}\hat A_{\alpha,{\bf
i},\sigma} \hat A^{\dagger}_{\alpha,{\bf i},\sigma}+U\hat P+
E_{g}^{III}. \label{ff1}
\end{eqnarray}

\emph{III.a Field induced localized ferromagnetism at $n=5/3$:} For
$N=5 N_c$ the ground state has the general form
\begin{equation}
|\Psi_{g}^{III}(5N_c,\delta)\rangle=\hat G^{\dagger}\hat
E^{\dagger}(\delta) |0\rangle \label{psiIII}
\end{equation}
where $\hat G^{\dagger}=\hat G^{\dagger}_{\uparrow}\hat
G^{\dagger}_{\downarrow}$, with $\hat
G^{\dagger}_{\sigma}=\prod_{\alpha=\pm}\prod_{{\bf i}} \hat
A^{\dagger}_{\alpha,{\bf i},\sigma}$, inserts two electrons with
spin $\sigma$ and two electrons with spin $-\sigma$ in each unit
cell. The first term of the Hamiltonian (\ref{ff1}) therefore
annihilates the ground state (\ref{psiIII}). In order to fulfill
also $\hat P\hat G^{\dagger}\hat E^{\dagger}(\delta)|0 \rangle=0$,
the operator $\hat E^{\dagger}(\delta)$ requires a special form,
which depends on the properties of $\hat G^{\dagger}$: For zero
flux ($\delta =0$) $\hat G^{\dagger}$ creates a double occupancy
on each site of the $s=3$ sublattice, where the diamonds touch,
thereby blocking any electron motion along the chain. The operator
$\hat E^{\dagger}(\delta=0)$, which adds one more electron to each
unit cell, can therefore place electrons only on sublattices
$s=1,2$, i.e., $\hat E^{\dagger}(0)= \prod_{{\bf i}} (\alpha_{\bf
i}\hat c^{\dagger}_{{\bf i}+{\bf r}_1,\sigma_{{\bf i},1}}+
\beta_{\bf i} \hat c^{\dagger}_{{\bf i}+{\bf r}_2,\sigma_{{\bf i},
2}})$ with arbitrary numerical coefficients $\alpha_{\bf i},
\beta_{\bf i}$ and spins $\sigma_{{\bf i},1},\sigma_{{\bf i},2}$.
Altogether the ground state is localized and non-magnetic with a
high spin degeneracy and has the form
\begin{eqnarray}
|\Psi_{g}^{III}(5N_c,0)\rangle = [\prod_{{\bf i},\sigma}\hat
c^{\dagger}_{{\bf i}, \sigma} (\hat c^{\dagger}_{{\bf i}+{\bf
r}_1,\sigma}+ \hat c^{\dagger}_{ {\bf i}+ {\bf r}_2,\sigma})] \hat
E^{\dagger}(0) |0\rangle.\,\, \label{ff2}
\end{eqnarray}
For finite flux ($\delta\ne 0$) $\hat G^{\dagger}$ no longer creates
a double occupancy on each site of the $s=3$ sublattice, thus
allowing electrons to move. To provide eigenvalue zero for the first
two terms of $\hat H$, (\ref{ff1}), $\hat E^{\dagger}(\delta\ne 0)$
must be chosen such that it introduces one electron with
\emph{fixed} spin (say, $\uparrow$) on each site of the $s=3$
sublattice. As a consequence there will be one $\uparrow$ spin
electron on all sites, i.e., these electrons are immobile
($N_{\uparrow}=3N_c,N_{\downarrow}=2N_c $). For $\delta\neq 0$ the
ground state for $N=5 N_c$ is therefore
a non-saturated ferromagnet with magnetization $M\propto (1-\Delta N/N_c)$
independent of $\delta$ and a finite one-particle localization
length $\xi_{\downarrow}$ (the proof proceeds as in
\cite{localization_length}).

\emph{III.b Field induced itinerant ferromagnetism and insulator-metal
transition at $5/3<n<2$:} The properties of the ground state at $\delta =0$
remain valid even for $N=5N_c+\Delta N$ with $1<\Delta N<N_c$ since the
ground state is now $|\Psi_g(N,\delta=0)\rangle=[\prod_{{\bf i}=1}^{\Delta
N}(\alpha'_{\bf i}\hat c^{\dagger}_{{\bf i}+{\bf r}_1,\sigma_{{\bf i},1}}+
\beta'_{\bf i}\hat c^{\dagger}_{{\bf i}+{\bf r}_2,\sigma_{{\bf i},2}})
|\Psi_g(5N_c,0) \rangle$. The product is
taken over $\Delta N$ arbitrary sites; the coefficients
$\alpha'_{\bf i}$, $\beta'_{\bf i}$ and spins $\sigma_{{\bf i},1}$,
$\sigma_{{\bf i},2}$ are arbitrary. Consequently, at $\delta=0$ one
obtains again a localized ground state.
For finite magnetic fields the ground state is instead
\begin{eqnarray}
|\Psi_{g}^{III}(N>5N_c, \delta ) \rangle =\prod_{\alpha=1}^{\Delta
N} \hat c^{\dagger}_{n_{\alpha},{\bf k}_{\alpha},-\sigma}
|\Psi_{g}^{III}(5N_c,\delta ) \rangle\, . \label{ff4}
\end{eqnarray}
Since the ground state now contains plane wave-type contributions
and $\delta\mu (5N_c+\Delta N)=0$ for $\Delta N>1$,
$|\Psi_{g}^{III}(N>5N_c, \delta ) \rangle$ corresponds to a
conducting, non-saturated ferromagnetic state with only mobile spin
$-\sigma$ electrons.
Hence the magnetic field induces an insulator-metal transition. In
the metallic state the net magnetic moment decreases linearly with
increasing density.

Solution III therefore has the following properties: At zero
magnetic field, $t_{\parallel}=0$, and a sublattice potential
$\epsilon=t_{\perp}- t^{-1}_{\perp}$, but otherwise arbitrary
$t_{\perp}<0$, $U>0$, it represents a localized non-magnetic ground
state over a continuous range of densities $n\geq5/3$. By contrast,
at finite magnetic field and the potential
$\epsilon=t_{\perp}/\cos\delta-\cos\delta/t_{\perp}$, but otherwise
for the same parameters, a non-saturated, ferromagnetic ground state
is obtained. This state is localized
at $n=5/3$, but gapless for $n>5/3$. For the latter density the majority
spin electrons are immobile and only the minority spin electrons are
itinerant. Therefore, by varying the magnetic field and the sublattice
potential one can tune from a localized, non-magnetic ground state in the
density range $n\geq5/3$ to a non-saturated ferromagnet, which is insulating
at $n=5/3$ and gapless for $n>5/3$.

In summary, by constructing exact ground states of electrons on the
diamond Hubbard chain in a magnetic field we showed that this
one-dimensional structure displays remarkably complex physical
properties which originate from flat single-electron bands. The
selected solutions describe flat-band
ferromagnetism, correlated half-metal behavior with spin-valve
features, and insulator-metal transitions. These properties do not
depend on the Zeeman interaction \cite{Zeeman}. The virtue of tuning
fundamentally different ground states through external magnetic
fields or site-selective potentials thereby points to new
possibilities for the design of electronic devices, which can switch
between insulating or conducting and ferromagnetic or non-magnetic
states.

We thank K. Byczuk for discussions. Support by the Hungarian
Scientific Research Fund through contract OTKA-T48782, the Alexander
von Humboldt Foundation, and the Deutsche Forschungsgemeinschaft
through SFB 484 is gratefully acknowledged.




\begin{references}
\bibitem{QHE} {\it The Quantum Hall Effect}, R. E. Prange and S. M. Girvin
(Eds.), Springer (New York, 1990).
\bibitem{Ramirez} R. Moessner and A. Ramirez, Physics Today {\bf 52}, 24
(2006).

\bibitem{int1}
H. Ishii, T. Nakayama, and J.I. Inoue, Phys. Rev. B{\bf 69}, 085325
(2004).

\bibitem{int1a}
H. Tamura, K. Shiraishi, T. Kimura, and H. Takayanagi, Phys. Rev.
B {\bf 65}, 085324 (2002).

\bibitem{int2}
R. Arita, Y. Suwa, K. Kuroki, and H. Aoki, Phys. Rev. Lett. {\bf 88},
127202 (2002); Y. Suwa, R. Arita, K. Kuroki, and H. Aoki, Phys. Rev. B
{\bf 68}, 174419 (2003).

\bibitem{int5}
J. Vidal, B. Doucot, R. Mosseri, and P. Butaud, Phys. Rev. Lett. {\bf
85}, 3906 (2000).

\bibitem{Jaksch}
D. Jaksch, C. Bruder, J. I. Cirac, C. W. Gardiner, and P. Zoller,
Phys. Rev. Lett. {\bf 81}, 3108 (1998).

\bibitem{Wu}
C. Wu, D. Bergman, L. Balents, and S. Das Sarma, preprint cond-mat/0701788.

\bibitem{Derzhko} O. Derzhko, A. Honecker, and J. Richter,
preprint cond-mat/0703295.

\bibitem{int6}
A. Mielke and H. Tasaki, Commun. Math. Phys. {\bf 158}, 341 (1993);
H. Tasaki, ibid. {\bf 242}, 445 (2003).

\bibitem{int7}
E. H. Lieb, Phys. Rev. Lett. {\bf 62}, 1201 (1989).

\bibitem{zv}
Z. Gul\'acsi and D. Vollhardt, Phys. Rev. Lett. {\bf 91}, 186401
(2003); Phys. Rev. B{\bf 72}, 075130 (2005).

\bibitem{non-canonical}
While the fermionic identity $(\hat A_{{\bf i},\sigma})^2=0$ holds,
the anti-commutator is non-canonical: $\{ \hat A_{{\bf i},\sigma}, \hat
A^{\dagger}_{{\bf j},\sigma} \} \ne \delta_{{\bf i},{\bf j}}$.

\bibitem{localization_length} This is proved by calculating
$\Gamma_{\bf r,-\sigma}$. For $|{\bf r}/{\bf a}|\gg 1$, $\Delta N=1$, $s=1$,
one finds $\Gamma_{{\bf r},-\sigma}/\Gamma^{(1)}_{\bf r,-\sigma}=1-[2+(1-
\cos ak)/t_{\perp}]^{-1}$. $\Gamma^{(1)}_{\bf r,-\sigma}$ is the plane wave
result and $k$ is the momentum of the electron added above $n=4/3$.

\bibitem{chemical_pot} Here $\delta\mu (N)=E_{g}^{II}(N+1)-2E_{g}^{II}(N)+
E_{g}^{II}(N-1)$ and $E_{g}^{II}(N)$ defined below Eq. (\ref{s4}).

\bibitem{Dieny} B. Dieny, Phys. Rev. B {\bf 43}, 1297 (1991).

\bibitem{DFT}
Density functional theory recently predicted also half-metallic
behavior for an organo-metallic wire; see V. V.
Maslyuk {\it et al.}, Phys. Rev. Lett. {\bf 97}, 097201 (2006).

\bibitem{Zeeman}
A Zeeman coupling
alters only solution I for $n<1/3$, where one would obtain a
ferromagnetic solution.

\end{references}
\end{document}